\documentclass[aps,twocolumn,superscriptaddress,altaffilletter,lengthcheck,
tightenlines,showpacs,showkeys]{revtex4}

\newcommand{\ben}{\begin{eqnarray}}
\newcommand{\een}{\end{eqnarray}}
\newcommand{\be}{\begin{equation}}
\newcommand{\ee}{\end{equation}}
\newcommand{\ba}{\begin{eqnarray}}
\newcommand{\ea}{\end{eqnarray}}
\newcommand{\n}{\label}

\newcommand{\ga}{\gamma}
\newcommand{\ro}{\rho}

\usepackage{graphicx}

\usepackage{epstopdf}

\usepackage[dvipdf]{epsfig}
\usepackage{color}
\usepackage{multirow}

\begin{document}
\title{Extended tachyon field using form invariance symmetry}

\author{Iv\'an E. S\'anchez G.}\email{isg.cos@gmail.com}
\affiliation{Departamento de Matem\'{a}tica, Facultad de Ciencias Exactas y Naturales, Universidad de Buenos Aires and IMAS, CONICET, Ciudad Universitaria, Pabell\'on I, 1428 Buenos Aires, Argentina}

\date{\today}
\bibliographystyle{plain}

\begin{abstract}
In this work we illustrate how form-invariance transformations (FIT) can be used to construct phantom and complementary tachyon cosmologies from standard tachyon field universes. We show how these transformations act on the Hubble expansion rate, the energy density, and pressure of the tachyon field. The FIT generate new cosmologies from a known ``seed'' one, in particular from the ordinary tachyon field we obtain two types of tachyon species, denominated phantom and complementary tachyon. We see that the FIT allow us to pass from a non-stable cosmology to a stable one and vice-versa, as appeared in the literature. Finally, as an example, we apply the transformations to a cosmological fluid with an inverse square potential, $V \propto \phi^{-2}$, and generate the extended tachyon field.


\end{abstract}
\vskip 1cm



\date{\today}
\bibliographystyle{plain}

\maketitle

\section{Introduction}

The tachyon field can play an important role in inflationary models \cite{Pad}-\cite{Jain} as well as in the present accelerated expansion, simulating the effect of the dark energy \cite{Pad}-\cite{Abramo}, \cite{Aguirre}-\cite{LuisTaq2} depending upon the form of the tachyon potential \cite{Pad}-\cite{Abramo}, \cite{Bagla}-\cite{Gilberto}. The tachyon is an unstable field which has becomes important in string theory through its role in the Dirac-Born-Infeld Lagrangian, because it is used to describe the D-brane action \cite{TaS}. It was shown that the tachyon field could play a useful role in cosmology independent of the fact that it can be an unstable field \cite{TaC}. Besides, it was pointed out in \cite {Pad} that the tachyon Lagrangian can be accommodated into a quintessence form when the derivates of the fields are small.

Several years ago, it was proposed that the tachyon Lagrangian could be extended in such a way to allow the barotropic index takes any value \cite{LuisTE} generating new species of tachyon called phantom and complementary tachyons in addition to the ordinary one \cite{phank}, \cite{LuisTaq2}, \cite{LuisTE}. The standard tachyon field can also describe a transition from an accelerated to a decelerated regime, behaving as an inflaton field at early times and as a matter field at late times. The complementary tachyon field always behaves as a matter field. The phantom tachyon field is characterized by a rapid expansion where its energy density increases with time \cite{LuisTaq2}, \cite{Shi}-\cite{Novos}.


On the other hand, form invariance transformations involve internal or external variables in such a way that the transformations preserve the form of the dynamical equations, i.e., they have a form invariance symmetry  (FIS) \cite{symm}. Particularly useful are the T-duality  \cite{Polchinsky} or ``scale-factor duality'' \cite{VET}.

A new kind of internal symmetry that preserves the form of the spatially flat Friedmann cosmology was found by one of the authors \cite{Luis}-\cite{upc}. There it was shown that the equations governing the evolution of FRW cosmologies have a FIS group. The FIT which preserves the form of those equations relates quantities of the fluid, energy density and pressure, with geometrical quantities such as the scale factor and Hubble expansion rate. The FIS introduce an alternative concept of equivalence between different physical problems meaning that essentially a set of cosmological models are equivalent when their dynamical equations are form invariant under the action of some internal symmetry  group \cite{FIS}.

The FIS makes possible to find exact solutions in several contexts and generate new cosmologies from a known ``seed'' one \cite{upc}-\cite{CM1}.

In this paper we will show that the FIT applied to the standard tachyon field, used as a seed, can generate the complementary tachyon field and the phantom tachyon field. Our main goal is to show that the extended tachyon field, is a consequence of the internal symmetry that preserves the form of the Einstein equations, in a FRW space-time. Additionally we will shown that the FIT allow us to pass from a non-stable cosmology to a stable one and vice-versa \cite{Gibb}, \cite{Barrow}. In particular we will analyze the tachyon field, driven by a potential depending inversely on the square of the scalar field. We will start with some seed cosmology and use FIT to obtain a different new one, for example passing form a accelerated to a super-accelerated scenario.

\section{FIS in flat FRW cosmology and linear FIT}

We will investigate an internal symmetry contained in the Einstein equations for a spatially flat FRW space-time
\be
\label{E1a}
3H^{2}=\rho\,
\ee
\be
\label{E1b}
\dot{\rho}+ 3H(\rho+p)=0\,
\ee
where $H=\dot a/a$ is the Hubble expansion rate and $a(t)$ is the scale factor. We assume that the universe is filled with a perfect fluid having  energy density $\rho$ and pressure $p$. The two independent Einstein equations have  three unknown quantities $(H, p, \rho)$, hence the system of equations (\ref{E1a})-(\ref{E1b}) has one degree of freedom. This allows to introduce FIT which involves those quantities,

\be
\n{tr}
\bar\rho=\bar\rho(\rho),
\ee
\be
\n{th}
\bar H=\left(\frac{\bar\rho}{\rho}\right)^{1/2}H,
\ee
\be
\n{tp+r}
\bar p+\bar\rho=\left(\frac{\rho}{\bar\rho}\right)^{1/2}\frac{d\bar\rho}{d\rho}\,\,(\rho+p).
\ee
Hence, the FIT (\ref{tr})-(\ref{tp+r}), generated by the invertible function $\bar\rho(\rho)$, make the job of preserving the form of the system of equations (\ref{E1a})-(\ref{E1b}) and the FRW cosmology has a FIS. The FIT (\ref{tr})-(\ref{tp+r}) map solutions of a define cosmology, through the variables $(H,p,\rho)$, into solutions of other system of equations,  defining  a different cosmology identified with the barred variables $(\bar{H}, \bar{p},\bar{\rho})$, forming a Lie group structure as is demonstrated in \cite{FIS}.

We present FIT induced by the linear generating function $\bar\ro=n^2\ro$ being $n$ a constant. After this choice Eqs. (\ref{tr})-(\ref{tp+r}) become
\be
\n{trn}
\bar\ro=n^2\ro,
\ee
\be
\n{thn}
\bar H=nH,\quad \Rightarrow \quad \bar a=a^n,
\ee
\be
\n{tpn}
(\bar\ro+\bar p)=n(\ro+p).
\ee
Hence, the linear transformation (\ref{trn}) leads to a linear combinations of the variables $\ro, H, p$ and a power transformation of the scale factor, obtained after having integrated $\bar H=nH$. Finally, the Eq. (\ref{tpn}) gives the transformation rule for the pressure of the fluid
\be
\n{tp}
\bar p=-n^2\ro+n(\ro+p).
\ee

In the case of considering two universes, each one of them filled with a perfect fluid for which we assume equations of state $\bar p=(\bar\ga-1)\bar\ro$ and $p=(\ga-1)\ro$ respectively,  the  barotropic index $\ga$ transforms as
\be
\n{tg}
\bar\ga=\frac{(\bar\ro+\bar p)}{\bar\ro}=\frac{\ro+p}{n\,\ro}=\frac{\ga}{n},
\ee
after using Eq. (\ref{trn}) along with Eq. (\ref{tpn}).

The existence of a Lie group structure opens the possibility of connecting the scale factor $a$ of a seed cosmology with the scale factor $\bar a =a^n$ of a different cosmology.

\section{The extended tachyon cosmology}

We turn our attention to the tachyon field and will show how it transforms under the FIT (\ref{trn})-(\ref{tpn}). We consider a scalar field $\phi$ of the tachyon-type with the self-interaction potential $V (\phi)$. The background energy density and pressure of the tachyon condensate, for a flat FRW cosmology, are
\be
\rho_\phi=\frac{V}{\sqrt{1-\dot{\phi}^2}}, \qquad \quad p_\phi=-V\sqrt{1-\dot{\phi}^2},  \label{roE}
\ee
respectively. The corresponding Einstein-Klein-Gordon (EKG) equations are
\be
\label{A}
3H^2=\frac{V}{\sqrt{1-\dot{\phi}^2}},
\ee
\be
\n{kg}
\ddot\phi+3H\dot\phi(1-\dot{\phi}^2)+\frac{1-\dot{\phi}^2}{V}\frac{dV}{d\phi}=0.
\ee

The equation of state for the tachyon is $p=(\gamma -1))\rho$, so the barotropic index is
\be
\n{gs}
\ga=\dot\phi^2.
\ee
with $0<\ga<1$ for Eqs. (\ref{roE}). The sound speed is $c^{2}_{s}=1-\ga>0$, and using (\ref{gs}), we can write
\be
\n{sos}
c^{2}_{s}=1-\dot\phi^2.
\ee

From equations (\ref{trn}) and (\ref{tp}), the transformed energy density and pressure of the tachyon field are given by
\be
\label{trt}
\bar{\ro}=\frac{\bar{V}}{\sqrt{1-\dot{\bar{\phi}}^2}}=\frac{n^{2}V}{\sqrt{1-\dot{\phi}^2}},
\ee
\be
\label{tpt}
\bar{p}=-\bar{V}\sqrt{1-\dot{\bar{\phi}}^2}=-\left(1-\frac{\dot{\phi}^2}{n}\right)\frac{n^{2}V}{\sqrt{1-\dot{\phi}^2}}.
\ee
So, we find that the tachyon field, the potential, the barotropic index and the sound speed transform linearly under the FIT (\ref{trn})-(\ref{tpn}),
\be
\label{tfv}
\dot{\bar{\phi}}^{2} =\frac{\dot{\phi}^{2}}{n}, \qquad \bar{V}=n^2 V \sqrt{\frac{1-\dot{\phi}^{2}/n}{1-\dot{\phi}^{2}}},
\ee
\be
\label{tgs}
\bar\ga=\frac{\ga}{n}, \qquad  \bar{c}^{2}_{s}=\frac{n-\dot\phi^2}{n}.
\ee
and the scalar field transforms as $\bar\phi=\phi/\sqrt{n}$.

We consider the Eqs. (\ref{roE}) with a barotropic index $0<\ga<1$ as a seed tachyon, called the ordinary tachyon. Using the first equation of (\ref{tgs}) we can get a barotropic index $\bar{\ga}=\ga/n<0$. Then, the energy density and the pressure of the barred cosmology are given by
\be
\label{trtp}
\bar{\ro}=\frac{\bar{V}}{\sqrt{1+\dot{\bar{\phi}}^2}}, \qquad  \bar{p}=-\bar{V}\sqrt{1+\dot{\bar{\phi}}^2},
\ee
These fluids represented by the Eq. (\ref{trtp}) with negative pressure and negative barotropic index describe phantom cosmologies. Moreover, we can get $1<\bar{\ga}=\ga/n$ under de condition $n<\ga$ applying the transformed rule (\ref{tgs}) to the barotropic index of the seed tachyon $0<\ga<1$. So, the energy density and the pressure of the barred fluid are
\be
\label{trtc}
\bar{\ro}=\frac{i\mid\bar{V}\mid}{i\sqrt{\dot{\bar{\phi}}^{2}-1}}=\frac{\mid\bar{V}\mid}{\sqrt{\dot{\bar{\phi}}^{2}-1}},
\ee
\be
\label{tptc}
\bar{p}=-i\mid\bar{V}\mid i\sqrt{\dot{\bar{\phi}}^2-1}=\mid\bar{V}\mid \sqrt{\dot{\bar{\phi}}^2-1},
\ee
while, these fluids described by the Eqs. (\ref{trtc}) and (\ref{tptc}) give rise to nonaccelerated expanding evolutions.

We used the ordinary tachyon field, Eqs. (\ref{roE}) with $0<\ga<1$, as a seed. With the application of the FIS Eqs. (\ref{trt})-(\ref{tgs}) we found the two species of tachyon fields, as in \cite{LuisTE}, the phantom tachyon Eqs. (\ref{trtp}) with a $\ga<0$ and the complementary tachyon Eqs. (\ref{trtc}) and (\ref{tptc}) with $1<\ga$. Therefore, the form invariance transformations allow us to extend the family of tachyon field.

Following Gibbons \cite{Gibb} and Barrow \textit{et. al.} \cite{Barrow}, the Einstein static universe containing a perfect fluid is always neutrally stable for the condition $c^{2}_{s}>1/5$. Therefore, the FIT Eq. (\ref{tgs}) allow us to pass from a non-stable cosmology to a stable one and vice-versa. For example, if we use a barotropic index $\ga_{0}=6/7$ as a seed solution with $c^{2}_{s}=1/7$, using the transformation rule Eq. (\ref{tgs}), we can get a stable cosmology with $c^{2}_{s}=5/7>1/5$ if $n=3$.

\subsection{Power-law expansion for the tachyon field}
Let us assume that the potential is an inverse square in terms of the tachyon field,
\be
\label{pot}
V(\phi)=\frac{V_0}{\phi^2},
\ee
with $V_0$ a constant. This potential, that diverges at $\phi=0$, fairly mimics the behavior of a typical potential in the condensate of bosonic string theory. The Eq. (\ref{pot}), leads to the power law expansion $a(t)=kt^{\delta}$, with $k$ a constant, if $\phi$ is the only source \cite{Pad} \cite{FE}. The tachyon field and the barotropic index are
\be
\label{CTqPL}
\phi=\left(\frac{2}{3\delta}\right)^{1/2}t,  \qquad  0<\gamma_{0}<1,
\ee
with
\be
\label{DTqPL}
\delta=\frac{1}{3}\left[1+\sqrt{1+4\beta}\right],  \qquad  \beta=\left(\frac{3V_{0}}{4}\right)^{2}.
\ee
For this reason the power law expansion appears to be a good example to illustrate how from a seed solution, characterized by particular values of the parameters $V_0$, $\ga_0$ and $k$, the FIS helps us to find the scalar field and the scale factor driven by inverse square potential (\ref{pot}) for any other value of those parameters. Applying the FIT (\ref{trn})-(\ref{tpn}) to the seed solution (\ref{CTqPL}), (\ref{DTqPL}) and using Eqs. (\ref{gs}), (\ref{tfv}) and (\ref{tgs}) we obtain the transformation rules for $V_0$ and $\ga_0$
\be
\label{ga0}
\bar{\ga_0}=\frac{\ga_0}{n},
\ee
\be
\label{V0}
\bar{V}_{0}= nV_{0}\sqrt{\frac{1-\frac{\ga_0}{n}}{1-\ga_0}}.
\ee

Therefore, the transformed tachyon field, for a barotropic index $\bar{\ga}<0$, is given by
\be
\label{plp}
\bar{\phi}=\left(\frac{2}{-3\mid\bar{\delta}\mid}\right)^{1/2}t,  \qquad  \bar{\delta}=\frac{1}{3}\left[1-\sqrt{1+4\bar{\beta}}\right].
\ee
These tachyon field solutions, Eq. (\ref{plp}), describe phantom cosmologies. Note that if $n=-1$ in Eqs. (\ref{ga0}) and (\ref{V0}) we can get the results of \cite{phank} for the phantom tachyon.

On the other hand, if the transformed barotropic index is $1<\bar{\ga}$, we get
\be
\label{plc}
\bar{\phi}=\left(\frac{2}{3\bar{\delta}}\right)^{1/2}t,  \qquad  \bar{\delta}=\frac{1}{3}\left[1\pm\sqrt{1-4\mid\bar{\beta}\mid}\right].
\ee
This type of tachyon field solution with $1<\ga_{0}$ is called the complementary tachyon solution, which represents stiff matter with a deceleration cosmology.

The scale factor $a(t)=kt^{\delta}$ transforms as $\bar{a}=a^{n}$, so the transformed scalar field is $\bar{a}=\bar{k}t^{\bar\delta}$ with $\bar{k}=k^{n}$ and $\bar{\delta}=n\delta$. The condition to has an inflation solution is that $1<\delta$ and it is represented by the solutions (\ref{CTqPL}). Notice that the exponent of the power law solution can takes positive or negative values provided that $n \in \Re$. We can see that this exponent is directly related with the barotropic index of the tachyon fluid, $\delta=2/3\ga_0$, therefore changing $n$ it is equivalent to allow that the $\bar{\gamma}_0$ varies over $\Re$. This simple fact leads us to a remarkable conclusion, there are new species of tachyons and the FIS has revealed their existence to us.

\section{conclusion}
As part of a long-term investigation \cite{Luis}-\cite{FIS} we have shown here that form invariance transformations can be used as tools for generating new solutions to the Einstein field equations, in this case the existence of two new kinds of extended tachyon fields were derived from the standard tachyon field ($0<\ga<1$): the complementary ($1<\ga$) and the phantom tachyon ($\ga<0$) fields, confirming the work made by one of the authors \cite{LuisTE}. In addition we see that the form invariance transformations allow us to pass from a neutrally unstable universe to a stable one \cite{Gibb}, \cite{Barrow}.

In particular, we have applied the method to obtaining phantom and complementary versions of FRW tachyon cosmologies, with an accent on power-law space-times generated by an inverse-square potential. We have found that the FIT transform the seed scale factor $a=kt^{\delta}$ into the power law solution $a=k^{n}t^{n\delta}$. For illustration purposes, if we start from a decelerated model with $2/3<\delta<1$, we can get a power-law inflation model with $\delta>1$ or a super-accelerated model (phantom model) with $\delta<0$. So, we have shown how FIT generate new cosmologies from a seed one.

\acknowledgments
This work was supported by Doctoral program of the Consejo Nacional de Investigaciones Cient\'{\i}ficas y T\' ecnicas (CONICET). I.S.G thanks IMAS, Math. Department, FyCEN-University of Buenos Aires. Also acknowledges Dr. Guillem Per\' ez-Nadal for useful discussions.



\vskip 1cm


\end{document}